\documentclass[a4paper,11pt]{article}
\usepackage{graphicx}
\usepackage{amsmath}

\input{tcilatex}

\begin{document}

\date{\today}

\begin{abstract}
Close insight into mathematical and conceptual structure of classical field
theories shows serious inconsistencies in their common basis. In other
words, we claim in this work to have come across two severe mathematical
blunders in the very foundations of theoretical hydrodynamics. One of the
defects concerns the traditional treatment of time derivatives in Eulerian
hydrodynamic description. The other one resides in the conventional
demonstration of the so-called Convection Theorem. Both approaches are
thought to be necessary for cross-verification of the standard differential
form of continuity equation. Any revision of these fundamental results might
have important implications for all classical field theories. Rigorous
reconsideration of time derivatives in Eulerian description shows that it
evokes Minkowski metric for any flow field domain without any previous
postulation. Mathematical approach is developed within the framework of
congruences for general 4-dimensional differentiable manifold and the final
result is formulated in form of a theorem. A modified version of the
Convection Theorem provides a necessary cross-verification for a
reconsidered differential form of continuity equation. Although the approach
is developed for one-component (scalar) flow field, it can be easily
generalized to any tensor field. Some possible implications for classical
electrodynamics are also explored.
\end{abstract}

{\large \textbf{On Two Complementary Types of Total Time Derivative in
Classical Field Theories and Maxwell's Equations }}

\bigskip \bigskip

$\;\;\;\;\;\;\;\;\;\;$\textbf{R. Smirnov-Rueda}

\bigskip

$\;\;\;\;\;\;\;\;\;\;$\textit{Applied Mathematics Department, }

\textit{$\;\;\;\;\;\;\;\;\;\;$Faculty of Mathematics }

\textit{$\;\;\;\;\;\;\;\;\;\;$Complutense University, }

\textit{$\;\;\;\;\;\;\;\;\;\;$28040 Madrid, Spain }

\bigskip




\section{Introduction}

This work treats some aspects of conceptual, logical and mathematical
structure of classical field theories. Put in other terms, we claim here to
have stumbled upon two severe mathematical blunders in the very foundations
of theoretical hydrodynamics as a corner-stone of all classical field
theories. More close insight shows that there are some difficulties in the
conventional approach to time derivatives. First indications of them could
already be found in Euler's seminal work "\textit{General Principles of the
Motion of Fluids}" (1755)$^{\cite{Euler}}$ whereas the other appeared in the
19th century in the demonstration of the so-called Convection Theorem. The
question about whether there are some reasons these defects resisted to be
seen up till now deserves special considerations elsewhere. However, we
shall make some comments in this respect in our attempt to clarify the
situation.

In modern retrospective we are aware of a certain unevenness in progressive
development of mathematics implying by it$^{\cite{morris}}$

\begin{quotation}
...\textit{false proofs, slips in reasoning, and inadvertent mistakes which
with more care could have been avoided. Such blunders there were aplenty.
The illogical development also involved inadequate understanding of
concepts, a failure to recognize all the principles of logic required, and
an inadequate rigor of proof; that is, intuition, physical arguments, and
appeal to geometrical diagrams had taken the place of logical arguments...}
\end{quotation}

In the first place, this uneasy state of affairs concerned mainly the
calculus, laid down by 16th and 17th-centuries scholars. However, the heroic
extension of the subject by 18th-century mathematicians (sometimes the 18th
century is called the heroic age in mathematics) to entirely new branches
(ordinary and partial differential equations, the calculus of variations,
differential geometry etc) did not imply a special effort in clarification
of basic concepts and logical justification of operations, frequently used
in the calculus. The problem of rigorization of the subject remained open.
According to the Morris Kline authorized opinion$^{\cite{morris}}$

\begin{quotation}
\textit{... the 18th century ended with the logic of the calculus and of the
branches of analysis built on the calculus in a totally confused state. In
fact, one could say that the state of the foundations was worse in 1800 than
in 1700. Giants, notably Euler and\textbf{\ }Lagrange, had given incorrect
logical foundations. Because these men were authorities, many of their
colleagues accepted and repeated uncritically what they proposed and even
built more analysis on their foundations...}
\end{quotation}

Even fundamental concepts of calculus proper such as a continuous function
and the derivative of a function were, to some extent, intuitive notions
well into the 19th century. By 1800 and early in the 19th century, among
mathematicians (Bolzano, Abel, Cauchy etc) certain concern was growing about
what was correct, inadequacies and contradictions in existing proofs,
confusions and vagueness in definitions etc. Cauchy, who started
clarification of foundations in the second decade of the 19th century and
exerted the most influence in rigorization of the calculus, decided to build
his approach on D'Alembert's limit concept. Notably, they were Bolzano and
Cauchy who gave a modern definition of the derivative$^{\cite{Kline}}$, i.e.
they defined the derivative as a limit, avoiding a purely formalistic
approach adopted earlier by Euler and Lagrange. Cauchy's work was followed
by many mathematicians and by Weierstrass, the most influent authority among
them, who removed all dependence on intuitive notions and by 1900 completed
the rigorization of the fundamentals of analysis.

At any rate, the requirement of rigor was also applicable to all branches of
applied mathematics and mathematical physics. Theoretical hydrodynamics was
no exception, although some aspects of specific empirical knowledge resisted
to be molded into adequate mathematical abstractions and remained on a level
of half-intuitive concepts for a longer period. It especially concerns the
notion of \textit{a fluid quantity} which found sound mathematical
clarification only at the end of the 19th century in works on geometrical
transformations (Klein, Lie etc). It is also worth reminding that the study
of functions, differential equations, basic notions of differential geometry
relevant to some extent in the framework of the theory of fluids and
elasticity was continued well in the 20th century. So that, perhaps it would
not be so groundlessly to think that some original defects might have
escaped a scrutiny of experts and could still take place in the conceptual,
logical or mathematical structure of classical field theories. However, a
deeper question about whether the installation of rigor led to relatively
major or minor corrections in mathematical foundations of hydrodynamics and
other field theories is, of course, a matter of additional investigation.

In what follows, our first intention is to fix basic conceptions of
mathematical hydrodynamics. We shall start with the physical background,
reminding the motivation for one or another description of fluids in
hydrodynamics. As a branch of mathematical physics, hydrodynamics deals with
a real physical world. Any physical description implies an observer who can
measure and compare measurable quantities. So that, a logically sound
description in hydrodynamics is inconceivable without an idealized observer.

In general, two complementary types of observers are thought to suffice in
order to provide a general description of the flow field kinematics or
dynamics. The first one (let us call it \textit{Lagrangian observer})
identifies individual elements or fluid particles and follows them along
their motion. This idea associates a non-zero hydrodynamic flow with a
non-zero geometrical transformation $H_{t}$ on the closure $\Omega _{0}$
such that the set $H_{t}\Omega _{0}$ represents the same individual bit of
fluid at time $t$. A position-vector of Lagrangian observer $\mathbf{r=r(}t%
\mathbf{)}$ coincides with the position-vector of an identified fluid
point-particle in a local coordinate system at rest. Therefore, the set of
coupled variables $\left\{ t,\mathbf{r(}t\mathbf{)}\right\} $ can be
naturally used for the mathematical description of fluid quantity $f$ from
the point of view of Lagrangian observer as a function of the type $f(t,%
\mathbf{r(}t\mathbf{))}$. This description is commonly known as \textit{%
Lagrangian specification} (or \textit{representation}) and is valid only if
the identification can be maintained by some kind of labeling usually
denoting the initial position $\mathbf{r}_{0}$ at instant $t_{0}$. Further
on, the set of coupled variables $\left\{ t,\mathbf{r(}t\mathbf{)}\right\} $
we shall refer sometimes as \textit{Lagrangian variables}.

The second one (let us call it \textit{Eulerian observer}) identifies a
fixed volume element or a fixed point of space in a local coordinate system
at rest. This observer conceives a description dissociated from
identification of individual bits of fluid. It makes use of the flow
quantity $f$ only as a function of local position $\mathbf{r}$ and time $t$,
i.e. as a function of the type $f(t,\mathbf{r)}$. Hence, implementation of
independent field (or \textit{Eulerian}) variables $\left\{ t,\mathbf{r}%
\right\} $ characterizes mathematical description of fluid quantities from
the point of view of Eulerian observer and represents what is commonly known
as \textit{Eulerian specification}. It has a special significance for
hydrodynamics and classical electrodynamics, since any practical attempt of
Lagrangian identification is unattainable in both cases. Obviously,
viewpoints of Lagrangian and Eulerian observers are complementary to each
other. Their convenience depends entirely on a particular context.

The next cardinal matter to be taken into consideration concerns the
description of time variations from the point of view of both observers. The
total time derivative for Lagrangian observer is expressed traditionally in
Eulerian field variables and is commonly known as Euler's directional (or
substantive) derivative. However, just the very treatment of the time
derivative for Eulerian observer looks like a rather delicate point in the
traditional approach. As we shall discuss it further, no equal standards of
rigor are applied in both cases. The rigorization of the conventional
formalism may have important implications for the differential form of
continuity equation as well as for some other aspects of classical field
theories and, especially, classical electrodynamics.


\section{Two complementary Lagrangian specifications of flow field}

As the first step, we shall fix basic concepts and notations of mathematical
hydrodynamics. We begin with the simplest assumptions for an ideal fluid
moving in a three-dimensional Euclidean domain. Let a mapping $H_{t}$
represent in Lagrangian description a geometrical transformation of the
initial closure $\Omega _{0}$ onto $H_{t}\Omega _{0}$ for the same
individual bit of fluid at time $t$. Then $H_{t}$ also represents the
function$^{\cite{meyer}}$:

\begin{equation}
\mathbf{r}=H_{t}\mathbf{r}_{0}=\mathbf{r(}t,\mathbf{r}_{0})  \label{odin}
\end{equation}
where points $\mathbf{r}$ and $\mathbf{r}_{0}$ denote the position-vector of
the fluid identifiable point-particle at time $t$ and initial time $t_{0}$,
respectively. The velocity of the particle along the trajectory is defined
as:

\begin{equation}
\mathbf{v=}\frac{d}{dt}\mathbf{\mathbf{r(}}t,\mathbf{\mathbf{r}}_{0}\mathbf{%
)=}\frac{\partial }{\partial t}\mathbf{r(}t,\mathbf{r}_{0})  \label{dva}
\end{equation}
where the initial position-vector $\mathbf{r}_{0}$ is assumed to be fixed or
time independent.

In the context of Eulerian description, \textit{a priori} there is no
identification and hence no explicit consideration of the function $\mathbf{r%
}=\mathbf{r(}t,\mathbf{r}_{0})$. The primary notion is the velocity field as
a function of position $\mathbf{r}$ in space and time $t$ on a fluid domain:

\begin{equation}
\frac{d\mathbf{r}}{dt}=\mathbf{v(}t,\mathbf{r})  \label{tri}
\end{equation}
where variables $\mathbf{r}$ and $t$ are originally uncoupled.

Picking up some initial point $\mathbf{r}_{0}=\mathbf{r(}t_{0})$, one
selects from a congruence (a set of integral curves of (\ref{tri})) a unique
solution. Thus, a formulation of the \textit{initial Cauchy problem}

\begin{equation}
\frac{d\mathbf{r}}{dt}=\mathbf{v(}t,\mathbf{r});\qquad \mathbf{r(}t_{0})=%
\mathbf{r}_{0}  \label{tria}
\end{equation}
is mathematically equivalent to an act of identification, allowing any
solution of (\ref{tria}) to be written in the form of (\ref{odin}). There is
a general consensus that this procedure can be taken as a rule for
translating from one specification to the other. However, more close insight
reveals a possibility of a different way of translation, alternative to the
traditional procedure. The treatment of time variations from the viewpoint
of Eulerian observer will show why it acquires special significance.

Let us explore this alternative. Our purpose here is to define a new
complementary Lagrangian specification by identifying a \textit{final}
closure instead of the initial one. We denote the final closure as $\Omega
_{f}$ and assume that it has always fixed shape and fixed position in space.
For further convenience, all our notation will be accompanied by \textit{%
tilde} when we refer to the Lagrangian description with identification of
the final closure $\tilde{\Omega}_{f}$. Then some geometrical transformation 
$\tilde{H}_{t}$ represents a mapping of the initial bit $\tilde{\Omega}_{0}$
at instant $t_{0}$ onto the final bit $\tilde{H}_{t}\tilde{\Omega}_{0}=%
\tilde{\Omega}_{f}$ at instant $t$. Since the closure $\tilde{\Omega}_{f}$
is assumed to be fixed in a local coordinate frame, the initial closure $%
\tilde{\Omega}_{0}$ becomes a function of space and time variables: $\tilde{%
\Omega}_{0}=\tilde{H}_{t}^{-1}\tilde{\Omega}_{f}$ (according to one of the
postulates of mathematical hydrodynamics$^{\cite{meyer}}$ for perfect
fluids, the transformation $H_{t}$ has the inverse $H_{t}^{-1}$, so there is
no obstacle in assuming the existence of the inverse mapping $\tilde{H}%
_{t}^{-1}$). In other words, the knowledge of the inverse geometrical
transformation $\tilde{H}_{t}^{-1}$ allows a reconstruction of initial bit $%
\tilde{\Omega}_{0}$ as well as its shape and space position at time $t_{0}$
from the knowledge of a fixed \textit{final} closure $\tilde{\Omega}_{f}$ at
time $t$.

When we deal with identifiable point-particle, $\tilde{H}_{t}$ represents
the function:

\begin{equation}
\mathbf{\tilde{r}}_{f}=\tilde{H}_{t}\mathbf{\tilde{r}}_{0}=\mathbf{\tilde{r}(%
}t,\mathbf{\tilde{r}}_{0})  \label{trib}
\end{equation}
which maps the initial position-vector $\mathbf{\tilde{r}}_{0}$ at time $%
t_{0}$ onto $\mathbf{\tilde{r}}_{f}$ at time $t$.

The whole situation can be seen schematically as follows: the initial
position of identifiable point-particle depends on a parameter $t$ and on a
fixed final position $\mathbf{\tilde{r}}_{f}$:

\begin{equation}
\mathbf{\tilde{r}}_{0}=\tilde{H}_{t}^{-1}\mathbf{\tilde{r}}_{f}=\mathbf{%
\tilde{r}}_{0}(t,\mathbf{\tilde{r}}_{f})  \label{tric}
\end{equation}
The longer is the lapse of time $t-t_{0}$, the larger is the distance which
a particle should travel through from $\mathbf{\tilde{r}}_{0}$ to $\mathbf{%
\tilde{r}}_{f}$.

It is obvious that in the new framework, Lagrangian specification can not be
linked to \textit{initial Cauchy problem} for (\ref{tri}). Only \textit{%
final Cauchy problem, }as we shall regard it, is appropriate for that
purpose:

\begin{equation}
\frac{d\mathbf{\tilde{r}}}{dt}=\mathbf{v(}t,\mathbf{\tilde{r}});\qquad 
\mathbf{\tilde{r}(}t,\mathbf{\tilde{r}}_{0})=\mathbf{\tilde{r}}_{f}
\label{trie}
\end{equation}
We remind here that the trajectory $\mathbf{\tilde{r}(}t,\mathbf{\tilde{r}}%
_{0})$ ends in the fixed point of space $\mathbf{\tilde{r}}_{f}$. In whole
similarity with the initial Cauchy problem, the\textit{\ final condition} $%
\mathbf{\tilde{r}(}t)=\mathbf{\tilde{r}}_{f}$ also selects a unique solution
from a congruence (a set of integral curves of the equation (\ref{tri})).
The set of coupled variables $\left\{ t,\mathbf{\tilde{r}(}t\mathbf{)}%
\right\} $ we shall refer as \textit{Lagrangian variables }for final Cauchy
problem.

To finish this Section, we conclude that both geometrical transformations $%
H_{t}$ and $\tilde{H}_{t}$ give Lagrangian description because they advance
any identified fluid point-particle from its position $\mathbf{r}_{0}$ (or $%
\mathbf{\tilde{r}}_{0}$) at time $t_{0}$ to its position at time $t$. In
both cases an imaginary observer \textit{follows the motion of the fluid} in
the direction of its velocity field $\mathbf{v}$. The only difference is
that one starts from a fixed initial point $\mathbf{r}_{0}$ and the other
ends in a fixed final point $\mathbf{\tilde{r}}_{f}$. In what follows, let
us explore some possible implementations of the transformation $\tilde{H}%
_{t} $.


\section{Fluid quantities and time derivatives for Lagrangian and Eulerian
observers}

To clarify our approach, all subsequent analysis will be based on a
consideration of one-component (scalar-field) ideal fluid moving in a
3-dimensional space closure. In general terms, we denote by fluid quantity
some regular function $f$ defined on a fluid domain. Any non-zero flow is
characterized by a non-zero velocity vector field $\mathbf{v}$ and \textit{%
vice versa}. Thus, in Lagrangian description the function $f$ is defined on
a set of Lagrangian variables $\left\{ t,\mathbf{r}(t)\right\} $, i.e. as a
function of the type $f(t,\mathbf{r}(t))$. If our approach is placed within
the traditional framework, then $\mathbf{r}(t)$ is an integral curve of the
equation (\ref{tria}) linked to the \textit{initial Cauchy problem}. In
other words, the transformation $H_{t}$:

\begin{equation}
\mathbf{r}_{0}=\mathbf{r}(t_{0})\quad \rightarrow \quad \mathbf{r(}t\mathbf{%
,r}_{0})=H_{t}\mathbf{r}_{0}  \label{chetyrea}
\end{equation}
defines the mapping of $f(t_{0},\mathbf{r}_{0})$ along the integral curve $%
\mathbf{r}(t)$ into a new function$^{\cite{dubrovin}}$:

\begin{equation}
f(t_{0},\mathbf{r}_{0})\quad \rightarrow \quad H_{t}f(t_{0},\mathbf{r}%
_{0})=f(H_{t}t_{0},H_{t}\mathbf{r}_{0})=f(t,\mathbf{r(}t,\mathbf{r}_{0}))
\label{chetyreb}
\end{equation}
where $H_{t}t_{0}=t$.

When the Lagrangian description is placed within the alternative framework
linked to the \textit{final Cauchy problem} (\ref{trie}), the geometrical
transformation $\tilde{H}_{t}$:

\begin{equation}
\mathbf{\tilde{r}}_{0}(t_{0})\quad \rightarrow \quad \mathbf{\tilde{r}(}t,%
\mathbf{\tilde{r}}_{0})=\tilde{H}_{t}\mathbf{\tilde{r}}_{0}=\mathbf{\tilde{r}%
}_{f}  \label{chetyrec}
\end{equation}
defines the mapping of $f(t_{0},\mathbf{\tilde{r}}_{0})$ along the integral
curve $\mathbf{\tilde{r}}(t)$ into a new function:

\begin{equation}
f(t_{0},\mathbf{\tilde{r}}_{0})\quad \rightarrow \quad \tilde{H}_{t}f(t_{0},%
\mathbf{\tilde{r}}_{0})=f(\tilde{H}_{t}t_{0},\tilde{H}_{t}\mathbf{\tilde{r}}%
_{0})=f(t,\mathbf{\tilde{r}(\tilde{r}}_{0},t))  \label{chetyred}
\end{equation}
where $\tilde{H}_{t}t_{0}=t$.

Both transformations $H_{t}$ and $\tilde{H}_{t}$ correspond to Lagrangian
description, i.e. represent the viewpoint of Lagrangian observer who follows
the motion of the fluid. Importantly, the geometrical interpretation based
on $H_{t}$ and $\tilde{H}_{t}$ (absent at the end of the 18th century) gives
a necessary \textit{clarification of the concept of fluid quantity}. Put in
qualitative terms, one can regard a \textit{fluid quantity} $f(t,\mathbf{r(}%
t))$ as a function value $f(t_{0},\mathbf{r}_{0})$ which is permanently
submitted to a non-zero geometrical transformation $H_{t}$, i.e. as a
function of the type $f(t,\mathbf{r(}t))=H_{t}f(t_{0},\mathbf{r}_{0})$ (note
again that $H_{t}$ is non-zero if and only if the velocity vector field $%
\mathbf{v}$ of fluid flow is non-zero).

In this context, the treatment of time variation of fluid quantities for
Lagrangian observer straightforwardly leads to the classical definition of
the total time derivative represented in Lagrangian variables:

\begin{equation}
\frac{d}{dt}f(t,\mathbf{r}(t))=\lim\limits_{t\rightarrow t_{0}}\frac{%
H_{t}f(t_{0},\mathbf{r}_{0})-f(t_{0},\mathbf{r}_{0})}{t-t_{0}}%
=\lim\limits_{t\rightarrow t_{0}}\frac{f(t,\mathbf{r}(t,\mathbf{r}%
_{0}))-f(t_{0},\mathbf{r}_{0})}{t-t_{0}}  \label{shest}
\end{equation}
as well as:

\begin{equation}
\frac{d}{dt}f(t,\mathbf{\tilde{r}}(t))=\lim\limits_{t\rightarrow t_{0}}\frac{%
\tilde{H}_{t}f(t_{0},\mathbf{\tilde{r}}_{0})-f(t_{0},\mathbf{\tilde{r}}_{0})%
}{t-t_{0}}=\lim\limits_{t\rightarrow t_{0}}\frac{f(t,\mathbf{\tilde{r}}(t,%
\mathbf{\tilde{r}}_{0}))-f(t_{0},\mathbf{\tilde{r}}_{0})}{t-t_{0}}
\label{shesta}
\end{equation}
Both definitions (\ref{shest}) and (\ref{shesta}) are equivalent so that
later on we shall refer only to the conventional expression (\ref{shest}).
Important to note that in this classical definition of the total time
derivative for Lagrangian observer, a function  $f$ is supposed to be
submitted to non-zero geometrical transformation $H_{t}$ associated with
non-zero flow velocity field $\mathbf{v}$.

The next natural step to do is to consider time variations in properly
Eulerian description (i.e. from the point of view of Eulerian observer). One
might anticipate here that formalistic Euler's approach had struck at this
point, relying on intuitive or half-intuitive concepts of the 18th century
calculus. As a consequence, Euler and his close followers interpreted a time
derivative of \textit{fluid quantities} for an observer at rest as a
standard partial time derivative:

\begin{equation}
\frac{\partial }{\partial t}f(t,\mathbf{r})=\lim\limits_{t\rightarrow t_{0}}%
\frac{f(t,\mathbf{r})-f(t_{0},\mathbf{r})}{t-t_{0}}  \label{shestb1}
\end{equation}
where $\mathbf{r}$ is a fixed point of space.

What is especially noteworthy about Euler's final result is that even after
the obvious progress in mathematical rigorization of the fluid quantity
concept had been achieved at the end of the 19th century, nobody seemed to
have worried about the following peculiarity. In the classical definition of
a partial time derivative (\ref{shestb1}), a function $f$ \textit{does not
explicitly possess mathematical characteristics of properly fluid quantity}
(recently, this fact was also critically pointed out in$^{\cite{smir}-\cite
{Chubyka}}$ but on different positions). Anyway, contrarily to what is
explicitly assumed in (\ref{shest}), in the definition (\ref{shestb1}) there
is no indication that $f$ is submitted to non-zero geometrical
transformation $H_{t}$ associated with non-zero flow velocity field $\mathbf{%
v}$. Therefore, (\ref{shestb1}) is not directly applicable to treat time
derivatives of fluid quantities.

To avoid traditionally formalistic approach (\ref{shestb1}), let us
carefully analyze time variations of fluid quantities in properly Eulerian
description. As a matter of fact, an observer should be placed at a fixed
position of space in order to undertake a study of any fluid quantity coming
through as a function of time. It can be easily achieved if one substitutes
a permanent monitoring of a fluid point-particle by a permanent
identification of a fixed point of space $\mathbf{\tilde{r}}_{f}$. Firstly,
Eulerian observer measures a fluid quantity $f(t_{0},\mathbf{\tilde{r}}_{f})$
carried by a point-particle on its way through $\mathbf{\tilde{r}}_{f}$ at
instant $t_{0}$. At time $t$, the previous particle has been replaced at $%
\mathbf{\tilde{r}}_{f}$ by another particle, carrying a fluid quantity $f(t,%
\mathbf{\tilde{r}}(t,\mathbf{\tilde{r}}_{0}))$, where $\mathbf{\tilde{r}}(t,%
\mathbf{\tilde{r}}_{0})=\mathbf{\tilde{r}}_{f}$. At time $t_{0}$ this
particle stayed at the point $\mathbf{\tilde{r}}_{0}$ different from $%
\mathbf{\tilde{r}}_{f}$. Thus, we realize here the necessity of the
practical implementation of the final Cauchy problem for a fluid
point-particle. This reasoning compels us to introduce a definition which
takes into account a non-zero geometrical transformation (associated with a
fluid flow) and which, therefore, differs from (\ref{shestb1}):

\begin{equation}
\frac{d^{\ast }}{d^{\ast }t}f(t,\mathbf{\tilde{r}}(t))=\lim\limits_{t%
\rightarrow t_{0}}\frac{\tilde{H}_{t}f(t_{0},\mathbf{\tilde{r}}_{0})-f(t_{0},%
\mathbf{\tilde{r}}_{f})}{t-t_{0}}=\lim\limits_{t\rightarrow t_{0}}\frac{f(t,%
\mathbf{\tilde{r}}_{f})-f(t_{0},\mathbf{\tilde{r}}_{f})}{t-t_{0}}
\label{shestb}
\end{equation}
where $\mathbf{\tilde{r}}_{f}=\mathbf{\tilde{r}}(t,\mathbf{\tilde{r}}_{0})$.

At first glance, there is no difference in symbolic notations between the
definition (\ref{shestb1}) and the right-hand side of the definition (\ref
{shestb}) so that they can be easily confused in a purely formalistic
approach when geometrical transformation $\tilde{H}_{t}$ is not taken into
account. Later on we shall prove that the mathematical object, defined by (%
\ref{shestb}), virtually differs from partial time derivative $\frac{%
\partial }{\partial t}$. To highlight this aspect in our discussions, we
shall denote the total time derivative in Eulerian description by upper 
\textit{asterisk}.

As the next step, let us consider analytical expressions for (\ref{shest})
and (\ref{shestb}) in Eulerian variables. The total time derivative is
conceived as a linear part of the rate of change of $f$ with respect to $t$.
Thus, when $t-t_{0}$ tends to zero, only linear part of geometrical
transformations $H_{t}$ makes sense for further discussions:

\begin{equation}
\mathbf{r}(t)=\mathbf{r}(t_{0})+(t-t_{0})(\frac{d\mathbf{r}}{dt}%
)_{t=t_{0}}+o(t-t_{0})=H_{t}\mathbf{r}(t_{0})  \label{shestc}
\end{equation}
where $\mathbf{r}_{0}=\mathbf{r}(t_{0})$ is the reference point of the
corresponding Taylor series; $(\frac{d\mathbf{r}}{dt})_{t=t_{0}}$ is the
initial velocity of a point-particle at $t_{0}$.

By analogy, the linear part of geometrical transformation $\tilde{H}_{t}$ is:

\begin{equation}
\mathbf{\tilde{r}}(t)=\mathbf{\tilde{r}}(t_{0})+(t-t_{0})(\frac{d\mathbf{%
\tilde{r}}}{dt})_{t=t_{0}}+o(t-t_{0})=\tilde{H}_{t}\mathbf{\tilde{r}}(t_{0})
\label{shestd}
\end{equation}
where $\mathbf{\tilde{r}}_{0}=\mathbf{\tilde{r}}(t_{0})$ is the reference
point of the corresponding Taylor series; $(\frac{d\mathbf{\tilde{r}}}{dt}%
)_{t=t_{0}}$ is evaluated at $\mathbf{\tilde{r}}_{0}$ different from $%
\mathbf{\tilde{r}}_{f}$ where Eulerian observer is placed. Importantly, for
further discussions we note that the reference point $\mathbf{r}_{0}$ in (%
\ref{shestc}) is fixed in a local coordinate system whereas the reference
point $\mathbf{\tilde{r}}_{0}=\tilde{H}_{t}^{-1}\mathbf{\tilde{r}}_{f}$ in (%
\ref{shestd}) is a function of a parameter $t$.

Applying (\ref{shestc}) to the definition (\ref{shest}), one gets:

\begin{equation}
\frac{d}{dt}f(t,\mathbf{r}(t))=\frac{\partial f}{\partial t}+\mathbf{v\cdot }%
\frac{\mathbf{\partial }f}{\partial \mathbf{r}}=\frac{Df}{Dt}  \label{sheste}
\end{equation}
where $\mathbf{v=}(\frac{d\mathbf{r}}{dt})_{t=t_{0}}$. The differential
operator $\frac{D}{Dt}=\frac{\partial }{\partial t}+\mathbf{v\cdot \nabla }$
is usually called \textit{substantive} or \textit{Euler's directional
derivative}.

Although this is well-known result, it needs some clarifying interpretation.
The left-hand side of (\ref{sheste}) is a symbolic expression for the total
time derivative for Lagrangian observer represented in Lagrangian variables
of the initial Cauchy problem. The right-hand side of (\ref{sheste})
represents the same result in Eulerian field variables. In fact, partial
derivatives $\frac{\partial f}{\partial t}$ and $\mathbf{\nabla }f$ make
sense only when variables $t$ and $\mathbf{r}$ are independent. Thus,
Euler's derivative $\frac{D}{Dt}$ allows a calculation of the total time
derivative meaningful for Lagrangian observer but does not require any
information on the particle trajectory $\mathbf{r}(t)$, indispensable in
Lagrangian specification. In other words, \textit{Euler's derivative} $\frac{%
D}{Dt}$ \textit{emulates the total time derivative of properly Lagrangian
description in infinitesimal vicinity of the initial point} $\mathbf{r}_{0}$%
. This interpretation is in agreement with what Euler himself thought two
and a half centuries ago and what is unreservedly accepted nowadays: the
directional derivative $\frac{D}{Dt} $ describes the rate of time variation
of material properties \textit{following the motion of the fluid}$^{\cite
{batchelor}}$.

Let us see what form will take in Euler's variables the total time
derivative (\ref{shestb}) considered by Eulerian observer that remains at
rest in a fixed point of space $\mathbf{\tilde{r}}_{f}$. Therefore, the set $%
\left\{ t,\mathbf{\tilde{r}}_{f}\right\} $ represents Eulerian variables for
(\ref{shestb}). It means that partial derivatives $\frac{\partial f}{%
\partial t}$, $\mathbf{\nabla }f$ and the value of fluid velocity $\mathbf{v}
$ have to be evaluated by Eulerian observer only locally at $\mathbf{\tilde{r%
}}_{f}$. This circumstance highlights the obvious inconvenience of the
reference point $\mathbf{\tilde{r}}_{0}$ which appears in the Taylor series (%
\ref{shestd}). In fact, any local reference system related to the reference
point $\mathbf{\tilde{r}}_{0}=\tilde{H}_{t}^{-1}\mathbf{\tilde{r}}_{f}$ is a
function of a time parameter $t$ (i.e. it is not a reference system at
rest). On the other hand, the reference system of Eulerian observer has to
be related to the fixed point of space $\mathbf{\tilde{r}}_{f}$. Hence, a
change of the reference point is required. It can be obtained by rewriting
the infinitesimal transformation (\ref{shestd}) as follows:

\begin{equation}
\mathbf{\tilde{r}}_{0}(t)=\mathbf{\tilde{r}}_{f}-(t-t_{0})(\frac{d\mathbf{%
\tilde{r}}}{dt})_{\mathbf{\tilde{r}}_{0}}+o(t-t_{0})  \label{sem}
\end{equation}
where $\mathbf{\tilde{r}}(t)=\mathbf{\tilde{r}}_{f}$ is a fixed point of
space.

Making use of the linearity of the transformation (\ref{shestd}), one can
conclude that a fluid point-particle arrives at $\mathbf{\tilde{r}}_{f}$ at
time $t$ with the velocity equal to the initial one $(\frac{d\mathbf{\tilde{r%
}}}{dt})_{\mathbf{\tilde{r}}_{0}}=(\frac{d\mathbf{\tilde{r}}}{dt})_{\mathbf{%
\tilde{r}}_{f}}$:

\begin{equation}
\mathbf{\tilde{r}}_{0}(t)=\mathbf{\tilde{r}}_{f}-(t-t_{0})(\frac{d\mathbf{%
\tilde{r}}}{dt})_{\mathbf{\tilde{r}}_{f}}+o(t-t_{0})  \label{sem1}
\end{equation}
where $\mathbf{\tilde{r}}_{f}$ is already the reference point. Let us denote
the transformation (\ref{sem1}) by $G_{t}$. Note that $G_{t}$ does not
change the arrow of time: $G_{t}t_{0}=t$.

Importantly, both transformations $G_{t}$ and $\tilde{H}_{t}$ are equivalent
only as infinitesimal transformations when higher order terms in
corresponding Taylor's series are not taken into account. Moreover, $G_{t}$
does not describe any mapping of function values along a congruence and is
necessary only to keep Eulerian observer at a fixed point of space $\mathbf{%
\tilde{r}}_{f}$, counteracting the flow drift. To grasp schematically the
underlying idea of (\ref{sem1}), one can fancy an observer on an escalator
(automatic staircase) running in a direction opposite to the direction of
the fluid velocity field $\mathbf{v}$ just to be always in the same point of
space $\mathbf{\tilde{r}}_{f}$.

Applying (\ref{sem1}) to the definition (\ref{shestb}), one gets the linear
part of the time variation of $f$ with respect to $t$ from the viewpoint of
Eulerian observer:

\begin{equation}
\frac{d^{\ast }}{d^{\ast }t}f(t,\mathbf{\tilde{r}}(t))=\frac{\partial f}{%
\partial t}-\mathbf{v\cdot }\frac{\mathbf{\partial }f}{\partial \mathbf{r}}=%
\frac{D^{\ast }f}{D^{\ast }t}  \label{sema}
\end{equation}
where $\mathbf{v}=\lim\limits_{t\rightarrow t_{0}}(\frac{d\mathbf{\tilde{r}}%
}{dt})_{\mathbf{\tilde{r}}_{f},t}=(\frac{d\mathbf{\tilde{r}}}{dt})_{\mathbf{%
\tilde{r}}_{f},t_{0}}$. In (\ref{sema}) we already use a common notation $%
\mathbf{r}$ for space variables instead of $\mathbf{\tilde{r}}_{f}$. This
change in notation is justified by the fact that $\mathbf{\tilde{r}}_{f}$
coincides with the space variable $\mathbf{r}$ of a local coordinate system
at rest. To make distinction between Euler's derivative $\frac{D}{Dt}$ and (%
\ref{sema}), we shall denote the latter by $\frac{D^{\ast }}{D^{\ast }t}$
and call \textit{local directional derivative}.

The interpretation of (\ref{sema}) is straightforward. The left-hand side is
a symbolic expression of the total time derivative for Eulerian observer
represented in Lagrangian variables of the final Cauchy problem, emulating
properly Eulerian description. The right-hand side represents the same
result in Euler's field variables. The schematized interpretation of the
transformation (\ref{sem1}) as counteraction of the flow drift helps to
grasp the meaning of the negative sign for the velocity value in (\ref{sema}%
).

Looking at (\ref{sheste}) and (\ref{sema}), one can see in explicit terms
the difference between total time derivatives for Lagrangian and Eulerian
observers. Further we shall test the correctness of the expression (\ref
{sema}) analyzing hydrodynamics conservation laws.


\section{Fluid quantities and time derivatives in 4-dimensional notation}

Before we proceed to the application of (\ref{sema}), it would be convenient
to make a certain generalization of the previous exposition. Let us choose a
4-dimensional, metric free framework for the description of an ideal fluid
in order to use the notion of \textit{Lie's derivative} as a particularly
important generalization of (\ref{shest}) on manifolds without metric. For
this purpose, we add a trivial statement $\frac{dt}{dt}=1$ to the
differential equation (\ref{tri}), in order to use a 4-dimensional notation:

\begin{equation}
\frac{dx^{i}}{dt}=V^{i}(x)  \label{semb}
\end{equation}
where $x=(x^{0},x^{1},x^{2},x^{3})=(t,\mathbf{r})$; $V=(1,\mathbf{v})$. To
satisfy the index conventions of modern differential geometry, upper indices
are used for coordinate functions $x^{i}(t)$, $i=0,1,2,3$. In further
discussions we shall also leave for the time variable $x^{0}$ its original
denomination $t$.

Let us formulate the initial Cauchy problem for (\ref{semb}):

\begin{equation}
\frac{dx^{i}}{dt}=V^{i}(x)\qquad \quad x_{0}=x(t_{0})  \label{vosem}
\end{equation}
where the lower index $0$ denotes the initial point on a manifold at $t_{0}$.

In full similarity with the previous Section, (\ref{vosem}) defines a
geometrical transformation $H_{t}$: $\ x_{0}\rightarrow x(t,x_{0})$ which
maps the initial point $x_{0}$ along the congruence onto $x(t,x_{0})$. Two
points $x_{0}$ and $x$ with parameters $t_{0}$ and $t$, respectively, are
related by the Taylor series:

\begin{equation}
x^{i}(t)=x^{i}(t_{0})+(t-t_{0})(\frac{dx^{i}}{dt})_{t_{0}}+\frac{%
(t-t_{0})^{2}}{2!}(\frac{d^{2}x^{i}}{dt^{2}})_{t_{0}}+...=H_{t}x^{i}(t_{0})
\label{vosema}
\end{equation}

The expression (\ref{vosema}) gives the finite motion $H_{t}x_{0}$ along the
integral curve whereas the first order (linear) operator $%
H_{t}x_{0}=x^{i}(t_{0})+(t-t_{0})(\frac{dx^{i}}{dt})_{t=t_{0}}$ gives only
an infinitesimal motion.

If a fluid quantity $f(x)$ is defined on the velocity vector field (\ref
{semb}), then the transformation $H_{t}$ describes the mapping of $f(x_{0})$
along the congruence into a new function $f(x)^{\cite{dubrovin}}$:

\begin{equation}
H_{t}f(x_{0})=f(H_{t}x_{0})=f(x)  \label{devyat}
\end{equation}

Thus, we arrive at the classical definition of the \textit{Lie derivative} $%
L_{V}$ along the vector field $V=(1,\mathbf{v)}$ written in the traditional
Lagrangian specification (i.e. linked to the initial Cauchy problem):

\begin{equation}
L_{V}\,f=[\frac{d}{dt}H_{t}\,f]_{t=t_{0}}=\lim\limits_{t\rightarrow t_{0}}%
\frac{1}{t-t_{0}}[f(x)-f(x_{0})]  \label{desyat}
\end{equation}
Note that (\ref{desyat}) gives a unique difference and therefore a unique
derivative.

When $t-t_{0}$ is too small, all higher order terms of the kind $\frac{t^{n}%
}{n!}(\frac{d^{n}}{dt^{n}})_{t=t_{0}}$ ($n\geq 2$) vanish in (\ref{vosema}),
and the mapping $H_{t}$: $\ x_{0}\rightarrow x(t,x_{0})$ is linear, holding
an explicit form:

\begin{equation}
H_{t}x_{0}=x_{0}^{i}+(t-t_{0})V^{i}(x_{0})+o(t-t_{0})  \label{odinadzat}
\end{equation}
The transformation (\ref{odinadzat}) gives the linear part of time
variation, providing analytic expression for Lie's derivative (\ref{desyat})
in Eulerian field variables:

\begin{equation}
L_{V}\,f=\frac{d}{dt}f(H_{t}\,x_{0})=V^{i}\frac{\partial f}{\partial x^{i}}
\label{dvenadzat}
\end{equation}
If we are in an ordinary Euclidean domain, $L_{V}$ takes a familiar form of
Euler's directional derivative$^{\cite{dubrovin}}$:

\begin{equation}
L_{V}\,f=V^{i}\frac{\partial f}{\partial x^{i}}=(\frac{\partial }{\partial t}%
+\mathbf{v\cdot \nabla )}f=\frac{Df}{Dt}  \label{trinadzat}
\end{equation}

Therefore, it is worth emphasizing here that the Lie derivative on a
differentiable manifold and its Euler's equivalent on an ordinary Euclidean
domain are traditionally defined entirely in the spirit of Lagrangian
description, i.e. when a point on a congruence or a fluid point-particle
remain permanently identified.

The infinitesimal transformation which in the previous Section we regarded
as $G_{t}$, takes the following equivalent form in 4-dimensional notation:

\begin{equation}
G_{t}x_{f}^{i}(t_{0})=x_{f}^{i}(t_{0})+(t-t_{0})V^{\ast i}(x_{f}(t_{0}))
\label{chet14}
\end{equation}
where $V^{\ast }=(1,-\mathbf{v})$ and the lower index $f$ denotes a fixed
space component $\mathbf{r}_{f}$. Note again that the transformation (\ref
{chet14}) does not describe any mapping along the congruence and makes sense
only as infinitesimal one. However, its use is fully justified, since it
effectively emulates the description of time variations perceived by
Eulerian observer. In 4-dimensional notation the infinitesimal
transformation $G_{t}$: $\ x_{f}(t_{0})\rightarrow G_{t}x_{f}(t_{0})$
implies the substitution of the function value $f(x_{f}(t_{0}))$ by $%
f(G_{t}x_{f}(t_{0}))$. In other words, it looks like some kind of\textit{\ }%
mapping of $f(x_{f}(t_{0}))$ into $f(G_{t}x_{f}(t_{0}))$:

\begin{equation}
G_{t}f(x_{f}(t_{0}))=f(G_{t}x_{f}(t_{0}))  \label{chet14a}
\end{equation}

This formulation allows us to adjust the framework of the conventional
definition (\ref{dvenadzat}) for our attempt to express analytically
Eulerian total time derivative on a 4-dimensional manifold as some kind of
Lie derivative on \textit{effective} local vector field $V^{\ast }$:

\begin{equation}
L_{V^{\ast }}\,f=[\frac{d}{dt}f(G_{t}\,x_{f}(t_{0}))]_{t=t_{0}}=V^{\ast i}%
\frac{\partial f}{\partial x^{i}}  \label{pyat15}
\end{equation}
Here we shall call $V^{\ast }=(1,-\mathbf{v})$ as \textit{effective velocity
vector field}, since it is meaningful for Eulerian observer in order to
counteract a flow drift.

In an ordinary Euclidean domain (\ref{pyat15}) takes the form of \textit{%
local directional derivative} established in the previous Section:

\begin{equation}
L_{V^{\ast }}\,f=V^{\ast i}\frac{\partial f}{\partial x^{i}}=(\frac{\partial 
}{\partial t}-\mathbf{v}\cdot \mathbf{\nabla )}f=\frac{D^{\ast }f}{D^{\ast }t%
}  \label{shest16}
\end{equation}

Let us remind that this result manifests the viewpoint of Eulerian observer,
i.e. it does not imply any sort of identification of points on a congruence
or\ fluid elements. Therefore, the directional derivative (\ref{pyat15})
defined on a general differentiable manifold and on an effective velocity
field $V^{\ast }$ should be regarded as a complementary counter-part of the
standard Lie derivative. The character of complementarity between (\ref
{dvenadzat}) and (\ref{pyat15}) is of the same nature as for the
relationship between their Euclidean analogies (\ref{trinadzat}) and (\ref
{shest16}) given in an ordinary Euclidean domain. Having in mind this
similarity with Lie's directional derivative, we can regard total time
derivatives $\frac{Df}{Dt}$ and $\frac{D^{\ast }f}{D^{\ast }t}$ for
Lagrangian and Eulerian observers, respectively, as \textit{complementary
4-dimensional directional derivatives} defined on 4-dimensional space-time
manifold.

Both types of directional derivatives $\frac{Df}{Dt}$ and $\frac{D^{\ast }f}{%
D^{\ast }t}$ can be analyzed in terms of 1-forms or real-valued functions of
vectors in 4-dimensional manifolds:

\begin{equation}
\omega =(\omega _{i})=(\frac{\partial f}{\partial x^{i}})  \label{tridzatdva}
\end{equation}
where $i=0,1,2,3$ and $(\frac{\partial f}{\partial x^{i}})=(\frac{\partial f%
}{\partial t},\mathbf{\nabla }f)$ in an ordinary Euclidean domain.

Now we point out that in tensor algebra the set $\{\omega _{i}V^{j}\}$ are
components of a linear operator or $( 
\begin{array}{c}
1 \\ 
1
\end{array}
)$ tensor. The formation of a scalar $\omega (V)$\ is called the contraction
of the 1-form $\omega $ with the vector $V$ and it is an alternative
representation of directional derivatives:

\begin{equation}
\frac{Df}{Dt}=\omega _{i}V^{i};\qquad \frac{D^{\ast }f}{D^{\ast }t}=\omega
_{i}V^{\ast i}  \label{tridzattri}
\end{equation}

The contraction of diagonal components of the tensor $\omega _{i}V^{j}$\ is
independent of the basis. Importantly, this law shows that both types of
directional derivatives $\frac{Df}{Dt}$ and $\frac{D^{\ast }f}{D^{\ast }t}$
are invariant and do not depend on a particular choice of a local coordinate
system. If there is a metric tensor defined on a manifold, then it maps
1-forms into vectors in a 1-1 manner. This pairing is usually written as:

\begin{equation}
\omega _{i}=g_{ij}\omega ^{j};\qquad V^{i}=g^{ij}V_{j}  \label{tridzat4}
\end{equation}
Therefore, from the point of view of tensor algebra, (\ref{tridzattri}) can
be represented as a scalar product in a 4-dimensional manifold with metric:

\begin{equation}
\frac{Df}{Dt}=g_{ii}\omega ^{i}V^{i};\qquad \frac{D^{\ast }f}{D^{\ast }t}%
=g_{ii}\omega ^{i}V^{\ast i}  \label{tridzat5}
\end{equation}
where $g_{ij}=\delta _{ij}$ is the Euclidean metric tensor. Important to
note that Euclidean metric appears here as an \textit{effective metric}
defined in infinitesimal 4-dimensional vicinity of ($t_{0}, \mathbf{r}_{0}$)
where $\mathbf{r}_{0}$ is the initial point from which Lagrangian observer
starts moving with the fluid.

A Minkowski metric is also consistently singled out for the local
directional derivative $\frac{D^{\ast }f}{D^{\ast }t}$:

\begin{equation}
\frac{D^{\ast }f}{D^{\ast }t}=g_{ii}\omega ^{i}V^{\ast i}=g_{ii}^{\ast
}\omega ^{i}V^{i}  \label{tridzat6}
\end{equation}
where $V^{\ast }=(1,-\mathbf{v})$; $g_{ij}^{\ast }=diag(1,-1,-1,-1)$ is
indefinite or Minkowski metric tensor. Similarly, Minkowski metric should be
understood here as an \textit{effective metric} defined in infinitesimal
4-dimensional vicinity of ($t_{0}, \mathbf{r}_{f}$) where $\mathbf{r}_{f}$
is the fixed point of space in which Eulerian observer is placed.

One of the advantages of the scalar product form is that it gives
orthonormal bases for space-time manifolds. For Lagrangian description, a
basis is Cartesian and a transformation matrix $\Lambda _{c}$ from one such
basis to another is orthogonal matrix:

\begin{equation}
\Lambda _{c}^{T}=\Lambda _{c}^{-1};\qquad ^{\prime }g_{ij}=\Lambda
_{c}^{-1}g_{ij}\Lambda _{c}  \label{tridzat7}
\end{equation}
These matrices $\Lambda _{c}$ form the symmetry group $O(4)$.

Likewise, for Eulerian description a Minkowski metric picks out a preferred
set of bases known as pseudo-Euclidean or Lorentz bases. A transformation
matrix $\Lambda _{L}$ from one Lorentz basis to another satisfies:

\begin{equation}
\Lambda _{L}^{T}=\Lambda _{L}^{-1};\qquad ^{\prime }g_{ij}^{\ast }=\Lambda
_{c}^{-1}g_{ij}^{\ast }\Lambda _{c}  \label{tridzat8}
\end{equation}
$\Lambda _{L}$ is called a Lorentz transformation and belongs to the Lorentz
group $L(4)$ or $O(3,1)$.

The point that needs to be emphasized here is the remarkable circumstance of
properly Eulerian description in evoking of the Minkowski metric \textit{%
without any previous postulation}. In other words, consistent mathematical
description of fluids is also perfectly compatible with the Lorentz symmetry
group. From the complementary standpoints of Lagrangian and Eulerian
observers it is clear that both kinds of total time (or 4-dimensional
directional) derivatives are valid only in their complementary contexts.
Hence, it also concerns the complementary relationship between Euclidean and
Minkowski metrics. In what follows we shall confine our attention on some
practical implications of (\ref{shest16}) in the classical field theory.


\section{Fluid contents in Lagrangian and Eulerian descriptions}

Let us now consider a fluid $f$-content in a $3$-dimensional space domain $V$%
, i.e a volume integral of the type $\int fdV$. Here we assume the function $%
f$ fulfils all standard conditions on integrability that allows us to
consider not only smooth or continuous integrand functions but a more
general class of fluid quantities with spatial discontinuities which may
take place in many practical examples (for instance, some extra particles of
dust moving in water introduce finite discontinuity into fluid density).
Nevertheless, regarding discontinuities, we shall restrict our approach only
by the simplest class or elementary discontinuities of finite size. Infinite
or delta-function types of discontinuities need additional suppositions on
integrability and, therefore, can be taken into consideration elsewhere on a
more rigorous basis.

In properly Lagrangian description $V$ is an identified macroscopic volume
domain moving with a fluid. If the bounding surface of a closure always
consists of the same fluid particles regardless any change of shape of the
volume $V$ then, as a result, no fluid flows through the volume surface.
Obviously, no time variation of a fluid content takes place and a
mathematical description can be used to express a conservation of $f$%
-content. If the above condition is not fulfilled and Lagrangian observer
identifies all fluid particles coming in and out the closure, there is a
general expression of the time variation, commonly known under the name of
the \textit{Convection Theorem}$^{\cite{meyer},\cite{batchelor}}$:

\begin{equation}
\frac{d}{dt}\int\limits_{V(t)}f(t,\mathbf{r}(t))dV=\int\limits_{V(t)}(\frac{%
Df}{Dt}+f\mathbf{\nabla }\cdot \mathbf{v})dV  \label{add1}
\end{equation}

In the left-hand side of (\ref{add1}) both volume $V(t)=H_{t}V(t_{0})$ and a
fluid quantity $f(t,\mathbf{r}(t))=H_{t}f(t_{0},\mathbf{r}_{0})$ are
permanently submitted to a non-zero geometrical transformation $H_{t}$, i.e.
they are given in Lagrangian specification linked to the initial Cauchy
problem whereas the right-hand side represents the same result in Eulerian
independent variables of the local reference system at rest. If there is no
flux through the surface, the Convection Theorem gives the conservation of
the fluid $f$-content. The restriction on conservation is usually written in
the standard differential form of continuity equation:

\begin{equation}
\frac{Df}{Dt}+f\mathbf{\nabla }\cdot \mathbf{v}=\frac{\partial f}{\partial t}%
+\mathbf{\nabla \cdot }f\mathbf{v}=0  \label{add2}
\end{equation}

The point of view of Eulerian observer is complementary. He picks out a
fixed 3-dimensional volume $V_{0}$ and studies a fluid $f$-content as a
function of time. Since the volume element is now fixed, the traditional
formalistic approach (see (\ref{shestb1}) in the Section 3) takes the time
derivative of Eulerian observer for the partial time derivative:

\begin{equation}
\frac{d}{dt}\int\limits_{V_{0}}fdV=\lim\limits_{t\rightarrow
t_{0}}\int\limits_{V_{0}}\left[ \frac{f(t,\mathbf{r})-f(t_{0},\mathbf{r})}{%
t-t_{0}}\right] dV=\int\limits_{V_{0}}\frac{\partial f}{\partial t}dV
\label{add3}
\end{equation}

In order to express the restriction on conservation when the time variation
of $f$-content is not zero, (\ref{add3}) should be equaled to the fluid
inflow or outflow through a bounding surface $\partial V_{0}$ of the volume $%
V_{0}$. It immediately leads to the integro-differential form of continuity
equation:

\begin{equation}
\int\limits_{V_{0}}\frac{\partial f}{\partial t}dV=-\int\limits_{\partial
V_{0}}f\mathbf{v\cdot }d\mathbf{S}=-\int\limits_{V_{0}}\mathbf{\nabla \cdot }%
f\mathbf{v}dV  \label{add4}
\end{equation}

The remarkable circumstance that the differential form of (\ref{add4})
coincides with (\ref{add2}) derived for the volume $V(t)$ in motion, is
traditionally associated with the cross-verification of the standard
differential form of continuity equation (\ref{add2}). However, from the
Section 3 we know that, generally speaking, partial time derivatives do not
represent time derivatives of fluid quantities perceived by Eulerian
observer. What attitude should we take then on the fact that the Convection
Theorem leads to the result obtained in the formalistic approach (\ref{add3}%
) which in no way refers to fluid quantities? In fact, we claim here to have
come across another mathematical blunder in the very demonstration of the
Convection Theorem. Elimination of defects implies mathematical
modifications in the conventional form of (\ref{add1}) (see Appendix A):

\begin{equation}
\frac{d}{dt}\int\limits_{V(t)}f(t,\mathbf{r}(t))dV=\int\limits_{V(t)}(\frac{%
\partial f}{\partial t}+f\mathbf{\nabla }\cdot \mathbf{v})dV  \label{add4a}
\end{equation}
where the set $\left\{ t,\mathbf{r}\right\} $ represents Eulerian field
variables associated with the local reference system at rest.

Before going further with implementation of the \textit{modified version of
the Convection Theorem} (\ref{add4a}), let us substitute (\ref{add3}) by a
rigorized description of the time derivative for Eulerian observer (see (\ref
{shestb}) in the Section 3). As a matter of fact, the terminology of the
Lagrangian description linked to the final Cauchy problem provides us with
the appropriate framework. Put in quantitative terms, Eulerian observer
measures a fluid quantity $f(t_{0},\mathbf{\tilde{r}}_{f})$ carried by a
point-particle on its way through $\mathbf{\tilde{r}}_{f}$ at instant $t_{0}$%
. At time $t$, the previous particle has been replaced at $\mathbf{\tilde{r}}%
_{f}$ by another particle, carrying a fluid quantity $f(t,\mathbf{\tilde{r}}%
(t,\mathbf{\tilde{r}}_{0}))$. If we apply this procedure to all fixed points
of space which form part of the fixed volume $V_{f}$, then we arrive at the
integral formulation:

\begin{equation}
\frac{d}{dt}\int\limits_{V_{f}}f(t,\mathbf{\tilde{r}}(t))dV=\lim\limits_{t%
\rightarrow t_{0}}\int\limits_{V_{f}}\left[ \frac{f(t,\mathbf{\tilde{r}}(t,%
\mathbf{\tilde{r}}_{0}))-f(t_{0},\mathbf{\tilde{r}}_{f})}{t-t_{0}}\right]
dV_{f}  \label{add5}
\end{equation}
where $\mathbf{\tilde{r}}(t,\mathbf{\tilde{r}}_{0})=$ $\mathbf{\tilde{r}}%
_{f} $ so that both values of $f$ represent the fluid property at the same
point of space $\mathbf{\tilde{r}}_{f}$.

The right-hand side of (\ref{add5}) makes use of Lagrangian variables of the
final Cauchy problem. If Eulerian variable are implemented, then according
to (\ref{sema}) (or (\ref{shest16})), the time variation of $f$-content for
Eulerian observer takes the following form:

\begin{equation}
\frac{d}{dt}[\int\limits_{V_{f}}f(t,\mathbf{\tilde{r}}(t))dV%
]_{t=t_{0}}=\int\limits_{V_{f}}(\frac{\partial f}{\partial t}-\mathbf{v}%
\cdot \mathbf{\nabla }f)dV  \label{add6}
\end{equation}
In our derivation of this general result we assumed that the gradient $%
\mathbf{\nabla }f$ is defined as a continuous function in all point of a
fluid domain.

To mark out the fact of complementarity between this Eulerian description
(for fixed volume domains) and the Lagrangian description (for volumes in
motion) regarded as the\textit{\ Convection Theorem} (\ref{add4a}), from now
on we shall call the result (\ref{add6}) as \textit{local Convection Theorem}
providing the following formal formulation:

\newtheorem{LCT}{Theorem} 
\begin{LCT}{(Local Convection Theorem):} Let $\mathbf{v}$ be a vector field
generating a fluid flow through a fixed 3-dimensional domain $V$ and
if $f(\mathbf{r},t)\in C^{1}(\bar{V})$, then

\begin{equation}
\frac{d}{dt}\int\limits_{V}fdV=\int%
\limits_{V}(\frac{\partial f}{\partial t}-\mathbf{v}\cdot
\mathbf{\nabla }f)dV \label{tridzat}
\end{equation}
where $dV$ denotes the fixed volume element.
\end{LCT}
Note that (\ref{tridzat}) is also applicable to arbitrary 1- and
2-dimensional closures of flow domains. Mathematical soundness of this
theorem can be easily seen on a simple 1-dimensional example considered in
Appendix B.

Since the volume domain $V_{f}$ is fixed in a local reference system at rest
and the flow vector field $\mathbf{v}$ is supposed to be non-zero, then a
time variation of $f$-content is unambiguously related to a flux of fluid
through the bounding surface $\partial V_{f}$. Thus, if (\ref{add6}) equals
the right-hand side of the equation (\ref{add4}), we obtain a modified
integro-differential version of $f$-content conservation law:

\begin{equation}
\int\limits_{V_{f}}(\frac{\partial f}{\partial t}-\mathbf{v}\cdot \mathbf{%
\nabla }f)dV=-\int\limits_{\partial V_{f}}f\mathbf{v}\cdot d\mathbf{S}%
=-\int\limits_{V_{f}}(\mathbf{\nabla \cdot }f\mathbf{v})dV  \label{add7}
\end{equation}

As integral form of the general conservation law, (\ref{add7}) should make
sense for continuous as well as for spatially discontinuous flows. In points
of spatial discontinuity of the function $f$, gradients $\mathbf{\nabla }f$
have singularities, hence invalidating the integrability properties of
integrands in both sides of the equation (\ref{add7}). This difficulty is
surmounted by itself when we realize that gradients $\mathbf{\nabla }f$,
which appear in both sides of (\ref{add7}), mutually cancel each other.
However, the problem remains open for the standard integro-differential form
of the conservation law. In fact, in the case of discontinuity of $f$, the
right-hand side of (\ref{add4}) contains a singularity which can not be
canceled.

If the volume $V_{f}$ in (\ref{add7}) tends to zero, we obtain \textit{a
modified differential form of continuity equation}:

\begin{equation}
\frac{\partial f}{\partial t}+f\mathbf{\nabla \cdot v}=0  \label{add8}
\end{equation}
Therefore, it is worthy to note that in the case of finite spatial
discontinuities of the function $f$, the partial differential equation (\ref
{add8}) does not possess untractable infinities related to the gradient $%
\mathbf{\nabla }f$. However, these singularities still take place in the
standard differential form of continuity equation (\ref{add2}).

The usefulness of the concept of conservation of fluid $f$-content comes
from its generality and also from its capability of cross-verifications of
the results obtained in complementary Lagrangian and Eulerian descriptions.
Thus, turning back to (\ref{add4a}), if there is no time variation, the
right-hand side equals to zero and in the left-hand side we arrive again at
the same conclusion (\ref{add8}) by means of the modified Convection
Theorem. The question about whether the previous cross-verification based on
ill-founded considerations (\ref{add1}) and (\ref{add3}) was just that
stumbling block which hindered the process of rigorization of foundations of
classical field theory, deserves some special clarifications elsewhere. Let
turn our attention on possible implications of the above-stated results for
classical electrodynamics.


\section{Local Convection Theorem and Maxwell's equations}

Another interesting task would be an application of the \textit{local
Convection Theorem} (\ref{tridzat}) to the integral form of Maxwell's
equations. To implement Eulerian description in hydrodynamics, one needs the
knowledge of fluid quantities and the velocity flow field as functions of
Eulerian variables in a local reference system at rest. The situation is
somewhat different in the case of classical electrodynamics. Evidently, $a$ $%
priori$ unknown nature of the velocity vector field for electromagnetic
field components cancels the validity of hydrodynamics common sense.
However, virtual applicability of the local Convection Theorem looks like
viable if, for instance, we restrict our approach to the consideration of
finite size charged particles moving with a constant velocity (if in the
limited case, the charge is of a delta-function type, then we suppose that
some additional conditions are added to the formulation of the local
Convection Theorem (\ref{tridzat}) in order to be applicable to this type of
charge density discontinuity).

In fact, Einstein's special relativity theory firmly established the
equivalence of inertial frames of reference in classical electrodynamics. If
a single electric charge is at rest in a local frame then its
electromagnetic field components do not explicitly depend on time from the
point of view of an observer in uniform motion. In other words, if we are in
the observer's inertial frame, electric field components will keep up
appearances with straight lines coming out of the charge source. Therefore,
if the charge velocity $\mathbf{v}_{q}$ is known, the velocity vector field $%
\mathbf{v=v}_{q}$ for components of electric and magnetic field is also
defined in the whole closure $V_{0}$.

Let us formulate Maxwell's equations in this particular case of one charge
system (microscopic version), using general notations. For Eulerian observer
all kind of space closures (volumes, bounding surfaces or curves) are fixed
in a local frame of reference at rest. The first pair of Maxwell's equations
does not refer to any time variation and describes the source of electric $%
\mathbf{E}$ and magnetic $\mathbf{H}$ field, respectively:

\begin{equation}
\int\limits_{S_{0}}\mathbf{E\cdot }d\mathbf{S}=\int\limits_{V_{0}}\mathbf{%
\nabla \cdot E}dV=4 \pi Q;\qquad \int\limits_{S_{0}}\mathbf{H\cdot }d\mathbf{%
S}=\int\limits_{V_{0}}\mathbf{\nabla \cdot H}dV=0  \label{sorok1a}
\end{equation}
where $V_{0}$ is a fixed volume; $S_{0}$ is a closed surface bounding $V_{0}$
and $Q$ is a whole electric charge inside $V_{0}$.

The differential formulation of (\ref{sorok1a}) is

\begin{equation}
\mathbf{\nabla \cdot E}=4 \pi \rho ;\qquad \mathbf{\nabla \cdot H}=0
\label{sorok1b}
\end{equation}
where $\rho $ is the charge density.

The second pair of Maxwell's integral equations refers to the time variation
of electric $\mathbf{E}$ and magnetic $\mathbf{H}$ field fluxes through a
fixed open surface $S_{0}$ bounded by a closed curve $C_{0}$:

\begin{equation}
\int\limits_{C_{0}}\mathbf{H}\cdot d\mathbf{l}=\frac{4\pi }{c}%
\int\limits_{S_{0}}\mathbf{j}\cdot d\mathbf{S}+\frac{1}{c}\frac{d}{dt}%
\int\limits_{S_{0}}\mathbf{E\cdot }d\mathbf{S}  \label{sorok2a}
\end{equation}

\begin{equation}
\int\limits_{C_{0}}\mathbf{E}\cdot d\mathbf{l}=-\frac{1}{c}\frac{d}{dt}%
\int\limits_{S_{0}}\mathbf{H\cdot }d\mathbf{S}  \label{sorok2b}
\end{equation}
where $\mathbf{j}=\rho \mathbf{v}$ is the\ charge current density.

Let us introduce a fluid quantity $f$ as a scalar product of two vectors $%
\mathbf{A}$ and $\mathbf{n}$, defined on the surface, i.e. $f=\mathbf{A\cdot
n}$, where $\mathbf{A}$ is some general vector field and $\mathbf{n}$ is a
unit vector normal to the surface. Since the local Convection Theorem (\ref
{tridzat}) is also meaningful for 1- and 2- dimensional flow domains, it can
be easily checked out that:

\begin{equation}
\frac{d}{dt}\int\limits_{S_{0}}(\mathbf{A\cdot n)}dS=\int\limits_{S_{0}}%
\left[ \frac{\partial \mathbf{A}}{\partial t}-(\mathbf{v\cdot \nabla )A}%
\right] \cdot \mathbf{n}dS=\int\limits_{S_{0}}\frac{D^{\ast }\mathbf{A}}{%
D^{\ast }t} \cdot d\mathbf{S}  \label{sorok2c}
\end{equation}

Despite of the limited validity of hydrodynamics approach to electromagnetic
field description, we make here an attempt to use (\ref{sorok2c}) in the
Eulerian representation of the second pair Maxwell's integral equations (\ref
{sorok2a}) and (\ref{sorok2b}) as follows:

\begin{equation}
\int\limits_{C_{0}}\mathbf{H}\cdot d\mathbf{l}=\int\limits_{S_{0}}\left[ 
\mathbf{\nabla ,H}\right] \cdot d\mathbf{S}=\frac{4\pi }{c}%
\int\limits_{S_{0}}\mathbf{j}\cdot d\mathbf{S}+\frac{1}{c}\int\limits_{S_{0}}%
\frac{D^{\ast }\mathbf{E}}{D^{\ast }t} \cdot d\mathbf{S}  \label{sorok4}
\end{equation}

\begin{equation}
\int\limits_{C_{0}}\mathbf{E}\cdot d\mathbf{l}=\int\limits_{S_{0}}\left[ 
\mathbf{\nabla ,E}\right] \cdot d\mathbf{S}=-\frac{1}{c}\int\limits_{S_{0}}%
\frac{D^{\ast }\mathbf{H}}{D^{\ast }t} \cdot d\mathbf{S}  \label{sorok5}
\end{equation}
or in a compact differential form:

\begin{equation}
\lbrack \mathbf{\nabla ,H]=}\frac{4\pi }{c}\mathbf{j+}\frac{1}{c}\frac{%
D^{\ast }\mathbf{E}}{D^{\ast }t};\qquad [\mathbf{\nabla ,E]=-}\frac{1}{c}%
\frac{D^{\ast }\mathbf{H}}{D^{\ast }t}  \label{sorok7b}
\end{equation}

If in this hydrodynamic formulation the velocity flow fields $\mathbf{v}$
for electromagnetic components $\mathbf{E}$ and $\mathbf{H}$ are known, then
the time derivative from the viewpoint of Eulerian observer is also defined
in explicit terms as $\frac{D^{\ast }}{D^{\ast }t}=\frac{\partial }{\partial
t}-\mathbf{v\cdot \nabla }$.

On the other hand, Maxwell's equations are compatible with the charge
conservation law. Since velocities of charge particles are measurable
values, their knowledge makes meaningful the direct application of the local
Convection Theorem (\ref{tridzat}) and, as a consequence, the law of the
conservation of $f$-content (\ref{add7}):

\begin{equation}
\int\limits_{V_{f}}(\frac{\partial \rho }{\partial t}-\mathbf{v}\cdot 
\mathbf{\nabla }\rho )dV=-\int\limits_{\partial V_{f}}\rho \mathbf{v}\cdot d%
\mathbf{S}=-\int\limits_{V_{f}}(\mathbf{\nabla \cdot }\rho \mathbf{v})dV
\label{sorok5a}
\end{equation}
or in modified differential form of continuity equation:

\begin{equation}
\frac{\partial \rho }{\partial t}+\rho\cdot \mathbf{\nabla }\mathbf{v} =0
\label{sorok5b}
\end{equation}

Let us write the differential form (\ref{sorok7b}) explicitly in terms of
partial derivatives and velocity vector field if they are presumably known:

\begin{equation}
\lbrack \mathbf{\nabla ,H]=}\frac{4\pi }{c}\mathbf{j+}\frac{1}{c}(\frac{%
\partial \mathbf{E}}{\partial t}-(\mathbf{v\cdot \nabla )E);\qquad }[\mathbf{%
\nabla ,E]=-}\frac{1}{c}(\frac{\partial \mathbf{H}}{\partial t}-(\mathbf{%
v\cdot \nabla )H)}  \label{32}
\end{equation}
Applying a general expression valid for any vector field $\mathbf{E}$ (or $%
\mathbf{H}$):

\begin{equation}
(\mathbf{v}\cdot \mathbf{\nabla })\mathbf{E}=\mathbf{v}(\mathbf{\nabla \cdot
E})-[\mathbf{\nabla },[\mathbf{v},\mathbf{E}]]  \label{sorok6}
\end{equation}
and having in mind the first pair of Maxwell's source equations (\ref
{sorok1b}) we arrive at:

\begin{equation}
(\mathbf{v}\cdot \mathbf{\nabla })\mathbf{E}=4 \pi\mathbf{j}-[\mathbf{\nabla 
},[\mathbf{v},\mathbf{E}]];\qquad (\mathbf{v}\cdot \mathbf{\nabla })\mathbf{H%
}=-[\mathbf{\nabla },[\mathbf{v},\mathbf{H}]]  \label{sorok6a}
\end{equation}
where $\mathbf{j}=\rho \mathbf{v}$.

It is worth reminding here that our approach to integral formulation of
Maxwell's equation has been originally submitted to a consideration of
uniformly moving charged particle. In this case the time dependence of
fields is implicit and all partial time derivatives vanish from (\ref{32}).
Substituting (\ref{sorok6a}) into (\ref{32}), we arrive at the
well-established relationship between quasistatic magnetic and electric
field strengths of uniformly moving charge or magnetic source, respectively$%
^{\cite{landau}}$:

\begin{equation}
\mathbf{H}=\frac{1}{c}[\mathbf{v},\mathbf{E}];\qquad \mathbf{E}=-\frac{1}{c}[%
\mathbf{v},\mathbf{H}]  \label{sorok7}
\end{equation}
where $\mathbf{v}$ is the velocity of a source.

Note that (\ref{sorok7}) makes sense only from the viewpoint of Eulerian
observer placed in the local inertial reference system at rest so that no
use of the special relativity relationships for field transformations has
been necessary.

It is very important to stress here that in our attempt to reconsider basic
differential equations, we leave without any modification the original
integral form of continuity as well as Maxwell's equations. Note that this
is the only form (not differential one) which had been verified by
experiments. So that no additional experimental confirmation of these
fundamental laws is implied in this approach. In fact, in this work we call
into question whether the conventional mathematical procedure of the
transition from integral equations with total time derivative for \textit{%
fluid quantities} to their differential form is correct. On the other hand,
the differential form of continuity as well as Maxwell's equations
constitutes the standard basis for providing mathematical solutions for the
classical electromagnetic theory. Hence what kind of differential equations
of classical electrodynamics have been under a scrutiny of experts for more
than a century?

The major fact that emerges from the above considerations is that there
appear two conflicting approaches to the differential form of Maxwell's
equations from the viewpoint of Eulerian observer. The traditional one is
indifferent to any specific manifestations of fluid quantities, treating the
time derivative for an observer at rest as a simple partial time derivative $%
\frac{\partial }{\partial t}$. The other starts out the same integral
formulation of Maxwell's equations but the time derivative for Eulerian
observer no longer coincides with the partial one. There appears an extra
term $\mathbf{v\cdot \nabla }$ which is one of the outcomes of the non-zero
flow velocity field, in what case it can be regarded as a convective term.

Therefore, the denial of the exclusive use of partial time derivatives to
describe time variation for an observer at rest and a recognition of a
deeper underlying meaning of the total time derivative in Eulerian
description imply inevitable changes in the structure of mathematical
solutions to Maxwell's equations (as regards this aspect, some alternative
frameworks for classical electrodynamics were recently discussed in $^{\cite
{smir},\cite{smirn}-\cite{chesonsm}}$). However, a wider analysis of the
integral form of Maxwell's equations on basis of the local Convection
Theorem does not look tractable at present stage. Perhaps, it is possible to
approach a description of electromagnetic field of a classical spinning
charged particle using above-considered approximations. Nevertheless, the
question about whether some postulates are indispensable in this and general
cases should be studied carefully elsewhere.


\section{Conclusions}

In this work we attempted to get a more detailed insight towards some
traditional aspects of mathematical and conceptual structure of theoretical
hydrodynamics. We found that no equal standards of rigor take place in
Lagrangian and Eulerian descriptions. The reconsidered account provides a
rigorous analytical approach to the treatment of time derivatives in
properly Eulerian description. To avoid traditional formalistic approach in
which the total time derivative of Eulerian observer is taken for the
partial time derivative, we realized the necessity of the practical
implementation of the final Cauchy problem for velocity field differential
equation. It justified a new definition for the total time derivative of
Eulerian observer which was regarded in this work also as the \textit{local
directional derivative.} The point that needs to be emphasized here is the
complementary character of the above introduced concept. It can be
considered as a complementary counter-part of the well-known Euler's
derivative.

By no means, the local directional derivative substitutes the Euler
mathematical construction. By contrary, it is shown that both types of total
time derivatives for Lagrangian and Eulerian observers (which could be
interpreted also as \textit{two complementary 4-dimensional directional
derivatives}) are equally valid but should be used in different contexts.

One of the interesting conclusions of the analytic expression for both
4-dimensional directional derivatives is that the choice between Lagrangian
and Eulerian types of flow field specification is equivalent to the choice
between space-time manifolds with Euclidean and Minkowski metric
respectively. Therefore, a consistent mathematical description of fluid
kinematics can be also compatible with the Lorentz group symmetry. Although
our approach was restricted by the consideration of one-component (scalar)
ideal flow field, the notion of the local directional derivative can be
easily generalized on Lie's derivatives for any general tensor field on
differentiable manifolds. Both types of Lie's derivative will correspond to
both complementary types of descriptions.

The concept of local directional derivative has been also applied to analyze
time variation of fluid contents in Eulerian description. The result has
been formulated in form of a theorem called here as the local Convection
Theorem meaningful for fixed space domains. Therefore, it should be refer as
complementary to the \textit{Convection Theorem} established for Lagrangian
description, i.e. for space domains moving with a fluid. Another unexpected
outcome of the approach developed in this work consists in modification of
the standard differential form of continuity equations. It also implies a
reconsideration of the differential form of Maxwell's equations since they
are compatible with the law of charge conservation, i.e continuity equation
for the charge density.

In place of concluding remark let us remind asserting and encouraging
attitude of a great mathematician. Gauss once wrote in his letter to Bessel
(quoted from$^{\cite{morris}}$)

\begin{quotation}
...\textit{One should never forget that the function [of complex variable],
like all mathematical constructions, are only our own creations, and that
when the definition with which one begins ceases to make sense, one should
not ask, what is, but what is convenient to assume in order that it remain
significant...}
\end{quotation}

\section{Acknowledgments}

The author thanks the referees for their valuable remarks.

\bigskip

\begin{center}
\textbf{APPENDIX A. Modified version of the Convection Theorem}
\end{center}

Let us consider a fluid $f$-content (which we shall denote by $F=\int fdV$)
when a macroscopic volume domain $V(t)$ is identified and is moving with the
fluid:

\begin{equation}
F(t)=\int\limits_{V(t)}f(t,\mathbf{r}(t))dV  \label{ap1}
\end{equation}

In the framework of traditional Lagrangian description linked to the initial
Cauchy problem, it implies the existence of non-zero geometrical
transformation $H_{t}$ (associated with non-zero velocity flow field $%
\mathbf{v}$):

\begin{equation}
V(t)=H_{t}V(t_{0});\qquad f(t,\mathbf{r}(t))=H_{t}f(t_{0},\mathbf{r}_{0})
\label{ap2}
\end{equation}

Since the time derivative of $F(t)$ is going to be treated here, we have to
consider only linear part of geometrical transformations $H_{t}$ when $%
t-t_{0}$ tends to zero:

\begin{equation}
\mathbf{r}(t)=H_{t}\mathbf{r}_{0}=\mathbf{r}_{0}+(t-t_{0})\mathbf{v}%
+o(t-t_{0})  \label{ap3}
\end{equation}
where $\mathbf{r}_{0}=\mathbf{r}(t_{0})$ belongs to the initial volume $%
V(t_{0})=V_{0}$ and $\mathbf{v=}(\frac{d\mathbf{r}}{dt})_{t=t_{0}}$ is a
local value of flow field velocity.

The geometrical transformation $H_{t}$ is algebraically represented by the
Jacobian determinant $\det \left\vert \frac{\partial H_{t}\mathbf{r}_{0}}{%
\partial \mathbf{r}_{0}}\right\vert $ which for the infinitesimal
transformation (\ref{ap3}) takes the following form$^{\cite{arnold}}$:

\begin{equation}
\det \left\vert \frac{\partial H_{t}\mathbf{r}_{0}}{\partial \mathbf{r}_{0}}%
\right\vert =1+(t-t_{0})\mathbf{\nabla \cdot v}+o(t-t_{0})  \label{ap3a}
\end{equation}

Thus, the evolution of a fluid content $F(t)$ can be written in original
variables $\mathbf{r}_{0}$ if the Jacobian has been specified for each value
of the parameter $t$:

\begin{equation}
F(t)=\int\limits_{V_{0}}f(t,\mathbf{r}_{0})\det \left\vert \frac{\partial
H_{t}\mathbf{r}_{0}}{\partial \mathbf{r}_{0}}\right\vert dV_{0}  \label{ap4}
\end{equation}

Let us now analyze the time derivative of a fluid $f$-content:

\begin{equation}
\frac{d}{dt}F(t)=\frac{d}{dt}\int\limits_{V_{0}}f(t,\mathbf{r}_{0})\det
\left\vert \frac{\partial H_{t}\mathbf{r}_{0}}{\partial \mathbf{r}_{0}}%
\right\vert dV_{0}  \label{ap5}
\end{equation}
or in the case of infinitesimal transformation:

\begin{equation}
\frac{d}{dt}F(t)=\frac{d}{dt}\int\limits_{V_{0}}f(t,\mathbf{r}_{0})\left[
1+(t-t_{0})\mathbf{\nabla \cdot v}+o(t-t_{0})\right] dV_{0}  \label{ap6}
\end{equation}

Importantly, in the representation (\ref{ap4}) of the integrand function $%
f(t,\mathbf{r}_{0})$ space variable $\mathbf{r}_{0}$ does not already depend
on time parameter $t$. Put in other terms, the expression (\ref{ap4})
implements the Eulerian independent variable $\mathbf{r}_{0}$ for the
integrand function $f(t,\mathbf{r}_{0})$ and the integration volume $V_{0}$
instead of the Lagrangian flow variable $\mathbf{r}(t)$ used in (\ref{ap1}).
The original time dependence of the function $f(t,\mathbf{r}(t))$ through
Lagrangian variable $\mathbf{r}(t)$ is now replaced by the time dependence
of the Jacobian determinant. Therefore, since $\mathbf{r}_{0}$ is assumed to
be time independent in (\ref{ap6}), the time derivative of $f(t,\mathbf{r}%
_{0})$ coincides with the partial time derivative according to the classical
definition:

\begin{equation}
\frac{d}{dt}f(t,\mathbf{r}_{0})=\frac{\partial }{\partial t}f(t,\mathbf{r}%
_{0})  \label{ap7}
\end{equation}

Thus we arrive at the \textit{modified version of the Convection Theorem}:

\begin{equation}
\frac{d}{dt}[F(t)]_{t=t_{0}}=\int\limits_{V(t)}(\frac{\partial f}{\partial t}%
+f\mathbf{\nabla }\cdot \mathbf{v})dV  \label{ap8}
\end{equation}
under the condition $t\rightarrow t_{0}$, i.e. $V(t)\rightarrow V_{0}$.

From these considerations it is now clear that the conventional approach
took the partial time derivative (\ref{ap7}) for Euler's directional
derivative $\frac{Df}{Dt}$.

\bigskip

\begin{center}
\textbf{APPENDIX B. Example of applicability of the Local Convection Theorem}
\end{center}

To highlight mathematical soundness of the \textit{local Convection Theorem}
(\ref{add6}) let us consider a simple example of 1-dimensional ideal flow
defined on a fixed 1-dimensional interval $[a,b]$ on $x$-axis. Let the set $%
\left\{ t,x\right\} $ be Eulerian variables and $f$ a regular function, for
instance:

\begin{equation}
f(t,x)=t+x^{3}  \label{app1}
\end{equation}
Additionally, the fluid flow is defined by the velocity vector field $%
\mathbf{v}$:

\begin{equation}
\frac{dx}{dt}=\mathbf{v}(t,x)=\frac{t}{x^{2}}  \label{app2}
\end{equation}

Let us choose some fixed point $\tilde{x}_{f}$ from the interval $[a,b]$ and
formulate the final Cauchy problem for the equation (\ref{app2}). It is
equivalent to the definition of Lagrangian variables $\left\{ t,\tilde{x}%
(t)\right\} $ for some identified point-particle:

\begin{equation}
\mathbf{v}(t,\tilde{x})=\frac{d\tilde{x}}{dt}=\frac{t}{\tilde{x}^{2}};\quad
\rightarrow \quad \int\limits_{\tilde{x}_{0}}^{\tilde{x}_{f}}\tilde{x}^{2}d%
\tilde{x}=\int\limits_{t_{0}}^{t}tdt  \label{app3}
\end{equation}
Note again that in any solution of the final Cauchy problem, the initial
starting point $\tilde{x}_{0}$ is a function of $t_{0},t$ and $\tilde{x}_{f}$%
:

\begin{equation}
\tilde{x}^{3}(t_{0},t,\tilde{x}_{f})=\tilde{x}_{f}^{3}+\frac{3}{2}t_{0}^{2}-%
\frac{3}{2}t^{2}  \label{app4}
\end{equation}

Let us now calculate separately the time variation of the fluid $f$-content
considered in both parts of the \textit{local Convection Theorem} (\ref{add6}%
). We first evaluate the left-hand side integral:

\begin{equation}
\int\limits_{a}^{b}f(t,\tilde{x}(t))d\tilde{x}_{f}=\int\limits_{a}^{b}[%
\tilde{x}_{f}^{3}+\frac{3}{2}t_{0}^{2}-\frac{3}{2}t^{2}+t]d\tilde{x}_{f}
\label{app5}
\end{equation}
then we have:

\begin{equation}
\int\limits_{a}^{b}f(t,\tilde{x}(t))d\tilde{x}_{f}=\frac{1}{4}(b^{4}-a^{4})+(%
\frac{3}{2}t_{0}^{2}-\frac{3}{2}t^{2}+t)(b-a)  \label{app6}
\end{equation}
as well as its time derivative:

\begin{equation}
\frac{d}{dt}\int\limits_{a}^{b}f(t,\tilde{x}(t))d\tilde{x}%
_{f}=(1-3t_{0})(b-a)  \label{app7}
\end{equation}
taken under the condition $t\rightarrow t_{0}$.

The right hand-side of (\ref{add6}) considers $f$-quantity in Eulerian
independent variables $\left\{ t,\tilde{x}_{f}\right\} $. Since $x_{f}$
coincides with space variable of the local $x$-axes at rest, we shall use
for it the common notation $x$ (i.e. without \textit{tilde)}. Partial
derivatives are calculated at $t_{0}$ in all fixed points $x$ from $[a,b]$:

\begin{equation}
\frac{\partial f}{\partial t}=1;\qquad \frac{\partial f}{\partial x}%
=3x^{2};\qquad \mathbf{v}=\frac{dx}{dt}=\frac{t_{0}}{x^{2}}  \label{app8}
\end{equation}
and we proceed to the evaluation of the following integral:

\begin{equation}
\int\limits_{a}^{b}(\frac{\partial f}{\partial t}-\mathbf{v}\frac{\partial f%
}{\partial x})dx=\int\limits_{a}^{b}(1-3t_{0})dx=(1-3t_{0})(b-a)
\label{app9}
\end{equation}

A simple comparison of (\ref{app7}) and (\ref{app9}) validates the
applicability of the local Convection Theorem to the ideal fluid flow
defined by (\ref{app1}) and (\ref{app2}):

\begin{equation}
\frac{d}{dt}[\int\limits_{a}^{b}f(t,x(t))dx]_{t=t_{0}}=\int\limits_{a}^{b}(%
\frac{\partial f}{\partial t}-\mathbf{v}\frac{\partial f}{\partial x})dx
\label{app10}
\end{equation}
where $\frac{dx}{dt}=\mathbf{v}(t_{0},x)$.



\begin{thebibliography}{99}
\bibitem{Euler}  L. Euler, \textit{Hist. de l'Acad. de Berlin}, \textbf{11}
274-315 (1755)

\bibitem{morris}  M. Kline, \textit{Mathematics: The Loss of Certainty }%
(Oxford University Press, New York, 1980)

\bibitem{Kline}  M. Kline, \textit{Mathematical Thought from Ancient to
Modern Times}, Vol. 2 (Oxford University Press, New York, 1972)

\bibitem{meyer}  R.E. Meyer, \textit{Introduction to Mathematical Fluid
Dynamics} (Wiley, 1972)

\bibitem{dubrovin}  B. Dubrovin, S. Novikov and A. Fomenko, \textit{Modern
Geometry}, Vol. 1 (Ed. Mir, Moscow, 1982)

\bibitem{smir}  A.E. Chubykalo and R. Smirnov-Rueda, \textit{Mod. Phys.
Lett. A}, \textbf{12}(1) 1-24 (1997)

\bibitem{Chub}  A.E. Chubykalo, R.A. Flores, J.A. Perez, \textit{Proceedings
of the International Congress}, 'Lorentz Group, CPT and Neutrino', Zacatecas
University (Mexico), 384 (1997)

\bibitem{Chubyk}  A.E. Chubykalo and R. Alvarado-Flores, \textit{Hadronic
Jour.}, \textbf{25} 159 (2002)

\bibitem{Chubyka}  A. Chubykalo, A. Espinoza and R. Flores-Alvarado, \textit{%
Hadronic Jour.}, \textbf{27}(6) 625 (2004)

\bibitem{batchelor}  G.K. Batchelor, \textit{Introduction to Fluid Dynamics}
(Cambridge University Press, Cambridge, 1967)

\bibitem{landau}  L.D. Landau and E.M. Lifshitz, \textit{Classical Theory of
Fields} (Nauka, Moscow, 1973)

\bibitem{smirn}  A.E. Chubykalo and R. Smirnov-Rueda, \textit{Phys. Rev. E}, 
\textbf{53}(5) 5373-5381 (1996)

\bibitem{smirno}  A.E. Chubykalo and R. Smirnov-Rueda, \textit{Phys. Rev. E}%
, \textbf{57}(3) 3683-3686 (1998)

\bibitem{smirnov}  R. Smirnov-Rueda, \textit{Found. Phys.}, \textbf{35}(1)
1-31 (2005)

\bibitem{chesonsm}  A. Chubykalo, A. Espinoza, V. Onoochin and R.
Smirnov-Rueda, Edts., \textit{Has the Last Word Been Said on Classical
Electrodynamics? New Horizons} (Rinton Press, Princeton, 2004)

\bibitem{arnold}  V.I. Arnold, \textit{Mathematical Methods of Classical
Mechanics} (Nauka, Moscow, 1974)
\end{thebibliography}
\end{document}